\documentclass[onecolumn, a4size, 11pt]{IEEEtran}
\usepackage{amsmath}
\usepackage{amssymb}
\usepackage{amsfonts}
\usepackage{graphicx}
\usepackage{epsfig}
\usepackage{subfigure}
\usepackage{psfrag}
\usepackage{cite}
\usepackage{url}

\title{Energy Efficiency Optimization for MIMO Broadcast Channels\footnote{
This work is supported by National Basic Research Program of China (973 Program)
2007CB310602.}\footnote{The source of this paper is partly presented in IEEE
Wireless Communications and Networking Conference (WCNC) 2012 \cite{EEWaterfillingDPC}.}
\footnote{The authors are with the Personal Communication Network \& Spread
Spectrum Laboratory, Department of Electrical Engineering and Information Science,
University of Science and Technology of China Hefei, Anhui, 230027,
China (email: suming@mail.ustc.edu.cn, lqiu@ustc.edu.cn).}
\footnote{Corresponding author: Ling Qiu, lqiu@ustc.edu.cn.}}

\author{Jie Xu and Ling Qiu}

\begin{document}
\maketitle \thispagestyle{empty}

\begin{abstract}
Characterizing the fundamental energy efficiency (EE) limits of MIMO broadcast channels
(BC) is significant for the development of green wireless communications. We address the
EE optimization problem for MIMO-BC in this paper and consider a practical power model, i.e.,
taking into account a transmit independent power which is related to the number of active
transmit antennas. Under this setup, we propose a new optimization approach, in which the transmit
covariance is optimized under fixed active transmit antenna sets, and then active transmit
antenna selection (ATAS) is utilized. During the transmit covariance optimization, we propose
a globally optimal energy efficient iterative water-filling scheme through solving a series
of concave fractional programs based on the block-coordinate ascent algorithm. After that,
ATAS is employed to determine the active transmit antenna set. Since activating more transmit
antennas can achieve higher sum-rate but at the cost of larger transmit independent power consumption,
there exists a tradeoff between the sum-rate gain and the power consumption. Here ATAS can
exploit the best tradeoff and thus further improve the EE. Optimal exhaustive search
and low-complexity norm based ATAS schemes are developed. Through simulations,
we discuss the effect of different parameters on the EE of the MIMO-BC.
\end{abstract}


\setlength{\baselineskip}{1.3\baselineskip}
\newtheorem{definition}{\underline{Definition}}[section]
\newtheorem{fact}{Fact}
\newtheorem{assumption}{Assumption}
\newtheorem{theorem}{\underline{Theorem}}[section]
\newtheorem{lemma}{\underline{Lemma}}[section]
\newtheorem{corollary}{\underline{Corollary}}[section]
\newtheorem{proposition}{\underline{Proposition}}[section]
\newtheorem{example}{\underline{Example}}[section]
\newtheorem{remark}{\underline{Remark}}[section]
\newtheorem{algorithm}{\underline{Algorithm}}[section]
\newcommand{\mv}[1]{\mbox{\boldmath{$ #1 $}}}

\section{Introduction}

Green or energy efficient wireless communications have drawn increasing attention these days. This is because of not only the exponential traffic growth with the
popularity of smart phones but also the limited energy source with ever higher prices. In order to achieve the most efficient energy usage for wireless communication systems, various innovative ``green'' technologies across different layers of protocol stacks are necessary \cite{YChenComMag}. Among other, how to maximize the bits-per-Joule energy efficiency (EE) is one of the major topics in the research of green wireless communications.

Meanwhile, downlink multiuser (MU) multiple-input multiple-output (MIMO) is becoming the key technology for the next generation cellular networks such as
long term evolution advanced (LTE-A) and worldwide interoperability for microwave access (WiMAX) due to its significant improvements on average data rate performance. To understand the theoretic data rate limits of downlink MU-MIMO, the capacity or spectral efficiency (SE) of MIMO broadcast channels (BC) has been studied in the literature, e.g. \cite{DPC,Duality,IterativeBC,DPC_Region}. It has been shown in \cite{DPC_Region} that the rate region achieved by dirty paper coding (DPC) is the capacity region of MIMO-BC. Under a sum transmit power constraint, efficient algorithms such as iterative water-filling \cite{IterativeBC,IterativeMAC} have been proposed based on convex optimization techniques to compute the maximum achievable sum-rate capacity of MIMO-BC. The development of these algorithms relies on the duality between MIMO-BC and MIMO multi-access channel (MAC) \cite{Duality}.

The EE of MIMO-BC is in general defined as the sum-rate of MIMO-BC divided by the total power consumption, which denotes the delivered bits per-unit energy measured in bits
per-Joule. In contrast to the research on SE of MIMO-BC which only considers transmit power constraints, studying the EE of MIMO-BC requires a comprehensive understanding on the power consumption of downlink MU-MIMO systems. In a typical cellular network, base stations (BS) take the main parts of power consumption. Regarding the power consumption of the BS, besides the transmit power, various power elements  of BS such as circuit, processing, cooling also account for significant portions of the total power. When the BS is deployed with multiple antennas, the total power consumption is highly related to the number of active transmit antennas, i.e., when an antenna is on, the corresponding active radio frequency (RF) chain consumes circuit, processing power etc..  Under this practical power model, the EE optimization of MIMO-BC has been rarely studied, since it is distinct from the SE optimization and is also non-trivial. Specifically, based on the practical power model, activating all transmit antennas and utilizing highest sum transmit power, which are always optimal for SE optimization, are not always optimal for EE optimization.

%

We first study the EE optimization of the MIMO-BC under a practical power model in this paper. We assume that the total power consumption of a BS consists of three parts. The first part is proportional to the sum transmit power, accounting for the power amplifier (PA) power, the second part is equal to a constant multiplying the number of active transmit antennas, accounting for the circuit and processing power of active RF chains, and the third part is a constant accounting for the baseband processing and cooling related power. Under this setup, the transmit covariance and active transmit antenna set should be jointly optimized. We propose a new optimization approach with transmit covariance optimization and active transmit antenna selection (ATAS) to maximize the EE of the MIMO-BC. At first, we find that the EE optimization problem under fixed active transmit antenna set is a concave fractional program, and propose an energy efficient iterative water-filling scheme to obtain the optimal covariance, which is proven to be globally optimal. After that, exhaustive search and norm-based ATAS are developed to determine the active transmit antenna set.

\subsection{Contributions}

We observe that the EE is affected by both the transmit covariances and the active transmit antenna set, and thus we propose  a new optimization approach with transmit covariance optimization and ATAS.
%

Under fixed active transmit antenna sets, we derive the optimal energy efficient transmit covariances at first. Employing the uplink-downlink duality, the nonconcave EE of MIMO-BC is transformed into a dual quasi-concave EE of MIMO-MAC. To solve the quasi-concave maximization problem in a well structured manner, we
separate it into three subproblems, i.e. unconstrained EE optimization, sum-rate maximization under sum transmit power constraint and sum transmit power minimization under sun-rate constraint. Since the latter two subproblems have been solved in the literature, we only need to address the unconstrained EE optimization problem. We propose a novel well structured energy efficient iterative water-filling scheme for the unconstrained EE optimization based on the
block-coordinate ascent algorithm. During each iteration, the transmit covariance optimization is formulated as a concave fractional program, which is solved through relating it to a parametric concave program and applying the Karush-Kuhn-Tucker (KKT) optimality conditions. Interestingly, the solution of each iteration has a feature of water-filling. We prove the convergence of the proposed energy efficient iterative water-filling and show that the proposed scheme has a much lower complexity than the standard interior-point methods.

A novel ATAS procedure is proposed to determine the optimal active transmit antenna set. Optimal exhaustive search and low-complexity norm-based selection schemes are both developed. A unique feature of the ATAS here is that after the selection, the inactive antennas should be switched off to save power, i.e., employing micro-sleep \cite{Micro_sleep} or discontinuous transmission (DTX) \cite{3GPP}. Simulation results show that ATAS can further improve the EE significantly and give insights about the effect of different system parameters on the EE. Nevertheless, during the implementation of ATAS, when some antennas are inactivated, it will be difficult to obtain the channel state information at transmitter (CSIT) for these antennas in the upcoming transmission slots. This is referred to as an {\it invisible CSIT problem}. We also discuss this problem's effect on the system design.

\subsection{Related Works}

There are a lot of literatures discussing the EE of point to point MIMO channels with transmit covariance optimization without antenna selection
\cite{Cui,chongCJ11,sprabhuenergyefficient,Kim2,EEPrecoding,AccurateEEMIMO}. The point to point MIMO channel can always be separated into
parallel sub-channels through singular value decomposition (SVD) or after MIMO detection. In this case, only power allocation across the sub-channels
needs to be optimized to maximize the EE \cite{chongCJ11,sprabhuenergyefficient,Kim2}. As the sub-channels are parallel, the solution is similar with the energy efficient power allocation in OFDM systems \cite{Miao3,sprabhuenergyefficient2}. Nevertheless, the optimization for point to point MIMO channels is not applicable for the MIMO-BC, as the MIMO-BC cannot be simply transformed into parallel sub-channels{\footnote{Employing zero-forcing (ZF) precoding can transform the MIMO-BC to non-interference sub-channels, which is however far from optimality and leaves much room for improvement in terms of spectral and energy efficiency \cite{DPC}.}}. There are few
literatures discussing the EE for the MIMO-BC. To the best of the authors' knowledge, only \cite{chong2011VTC} and our previous work \cite{Xu1}
addressed the EE of the MIMO-BC, but they both assumed linear precoding design and equal transmit power allocation for simplicity. These assumptions make both works far away from the optimal solution.  To optimize the EE of the MIMO-BC, deriving the well structured transmit covariances   is a challenge.

Antenna selection is a widely discussed technology in spectral efficient MIMO systems, both at the transmitter and receiver side, e.g. in
\cite{AntennaSelectionOverview,TXRXAntennaSelection,TXSelBD,RXSelBD,MultiModeBD}. However, the spectral efficient transmitter antenna selection
\cite{AntennaSelectionOverview,TXRXAntennaSelection,TXSelBD} is always performed to choose the active antennas when the number of RF chains is smaller than the number of antennas. Meanwhile, the receive antenna selection \cite{RXSelBD,MultiModeBD} is always performed jointly with the ZF precoding to approach the asymptotic optimal performance. In our scenario, as DPC is employed, the receive antenna selection is not required. Moreover, as we consider the case when the number of transmit antennas is equal to RF chains, the purpose of ATAS is to save power through turning off the inactive RF chains, and thus the exhaustive search and norm-based ATAS schemes are different from the spectral efficient antenna selection. Furthermore, there exists another challenge of invisible CSIT problem, which is also discussed in this paper.

\subsection{Organization and Notation}

The rest of this paper is organized as follows. Section \ref{Sec2} introduces the system model and problem formulation. Section \ref{Sec3}
proposes the transmit covariance optimization to maximize the EE under fixed active transmit antenna set. Section \ref{Sec4} proposes the energy efficient ATAS and discusses the
implementation issue in the realistic systems. The simulation results and discussions are given in Section \ref{Sec5}. Finally, section \ref{Sec6} concludes this paper.

Regarding the notation, bold face letters refer to vectors (lower case) or matrices (upper case). The superscript $H$ and $T$ represent the
conjugate transpose and transpose operation, respectively. ${\rm Tr}(\cdot)$ denotes the trace of the matrix.

\section{System Model and Problem Formulation}\label{Sec2}

\begin{figure}[t]
\begin{center}
\includegraphics[width=13cm] {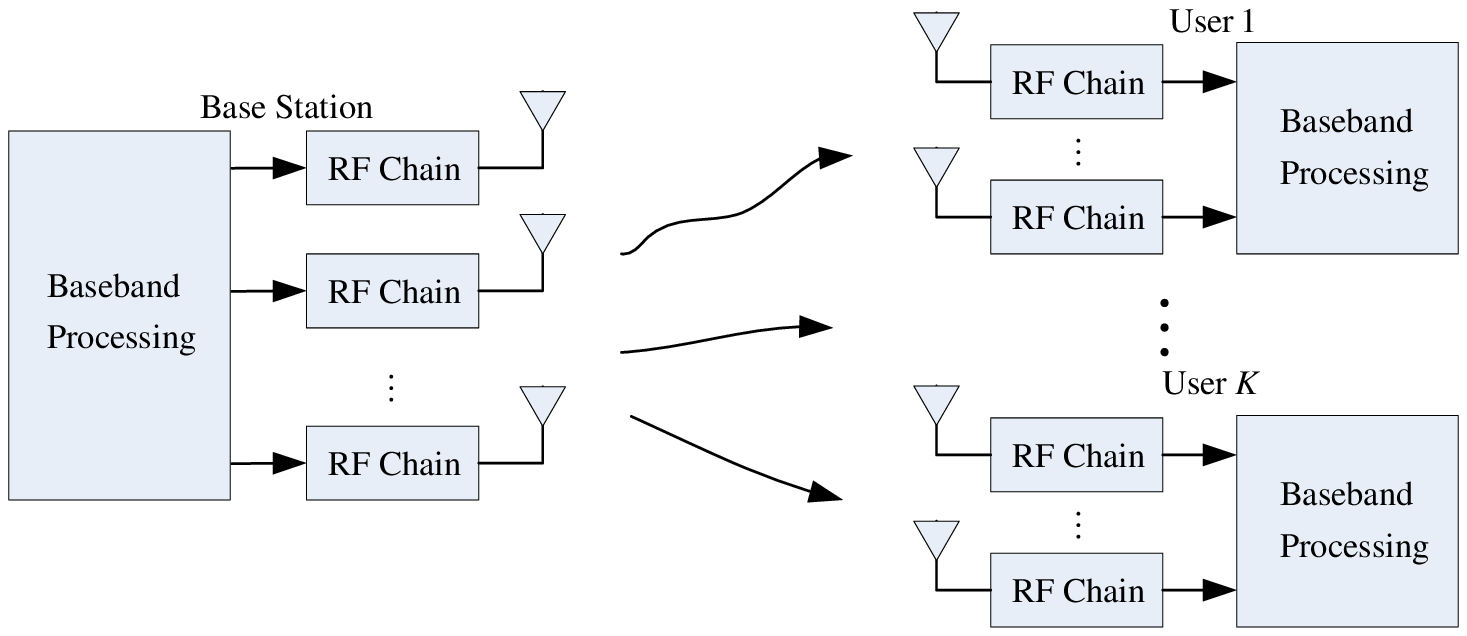}
\end{center}
\caption{System model of the MIMO-BC.} \label{fig0000}
\end{figure}

The system consists of a single BS with $M$ antennas and $K$ users each with $N$ antennas, which is shown in
Fig. \ref{fig0000}. We assume that the number of RF chains is equal to the number of antennas{\footnote{The results here can be extended to
the general case with different antenna number at each user and are also applicable to the multi-cell scenario with BS cooperation. Moreover, if
the number of RF chains is smaller than the antennas, our results can be simply extended after some modifications of the ATAS.}}. Denote the channel matrix from the BS to all users as ${\bf{H}} \in {\mathbb{C}}^{NK \times M}$ with ${\bf{H}} = [{\bf{H}}_1^T, {\bf{H}}_2^T, \ldots, {\bf{H}}_K^T ]^T$, where ${\bf{H}}_i \in {\mathbb{C}}^{N \times M}$ is the channel matrix from the BS to the $i$th user. As the number of active transmit antennas at the BS has a significant impact on the EE, selecting the active transmit antennas is important. We consider that the selected active transmit antenna set is ${\mathcal T} \subseteq \{1,\ldots,M\}$ with the number of active transmit antennas $M_{a} = |{\mathcal T}|$, and denote the channel matrix from the BS's active
transmit antennas to the users as ${\bf{H}}_{{\mathcal T}} \in {\mathbb{C}}^{NK \times M_a}$, with ${\bf{H}}_{\mathcal T} = [{\bf{H}}_{{\mathcal T},1}^T, {\bf{H}}_{{\mathcal T},2}^T, \ldots, {\bf{H}}_{{\mathcal T},K}^T ]^T$, where ${\bf{H}}_{{\mathcal T},i} \in {\mathbb{C}}^{N \times M_a}$ is the channel matrix from the BS's active transmit antennas to the $i$th user.

The downlink channel can be denoted as
\begin{equation} \label{eq1}
\begin{array}{l}
{\bf{y}}_i = {\bf{H}}_{{\mathcal T},i} {\bf{x}} + {\bf{n}}_i, i=1,\ldots,K,
\end{array}
\end{equation}
where ${\bf{n}}_i\in {\mathbb{C}}^{N \times 1}$ is the independent Gaussian noise with each entry ${\mathcal{CN}}(0,\sigma^2)$, ${\bf{x}} \in {\mathbb{C}}^{M_a \times 1}$ is the transmitted signal on the downlink. Meanwhile, ${\bf{x}} = {\bf{x}}_1 + \ldots + {\bf{x}}_K$ where ${\bf{x}}_i$ is the transmitted signal for user $i$ and ${{\bf{\Sigma }}_{\mathcal T,i}} = {\mathbb E}\left( {\bf{x}}_i {\bf{x}}_i^H\right)$ is the transmit covariance matrix for user $i$. Frequency flat fading channels with bandwidth $W$ are considered. Channel state information (CSI) is assumed to be perfectly known at the transmitter and receivers. CSIT can be acquired through uplink feedback in the frequency division duplex (FDD) systems or through uplink channel estimation in the time division duplex (TDD) systems. Although uplink feedback and channel estimation would induce imperfect CSIT, our optimization principle can be extended to that imperfect CSIT case based on the framework of robust optimization.

Given transmit covariances ${{\bf{\Sigma }}_{\mathcal T,i}}, i=1,\ldots,K$, the sum-rate of the MIMO-BC achieved by DPC is given by
\begin{align} \label{eq:sysmodel:1}
{C_{{\rm{BC}}}}\left( {{\bf{H}}_{\mathcal T,1}},\ldots,{{\bf{H}}_{\mathcal T,K}}, {{\bf{\Sigma }}_{\mathcal T,1}},\ldots,{{\bf{\Sigma }}_{\mathcal T,K}}  \right)
 = &  W\log \left| {{\bf{I}} + \frac{1}{\sigma^2}{{\bf{H}}_{\mathcal T,1}}{{\bf{\Sigma }}_{\mathcal T,1}}{\bf{H}}_{\mathcal T,1}^H} \right| \nonumber\\ &+
  W \log \frac{{\left| {{\bf{I}} + \frac{1}{\sigma^2}{{\bf{H}}_{\mathcal T,2}}\left( {{{\bf{\Sigma }}_{\mathcal T,1}} + {{\bf{\Sigma }}_{\mathcal T,2}}} \right){\bf{H}}_{\mathcal T,2}^H} \right|}}{{\left| {{\bf{I}} + \frac{1}{\sigma^2}{{\bf{H}}_{\mathcal T,2}}\left( {{{\bf{\Sigma }}_{\mathcal T,1}}} \right){\bf{H}}_{\mathcal T,2}^H} \right|}}+  \cdots\nonumber\\ &
 + W\log \frac{{\left| {{\bf{I}} + \frac{1}{\sigma^2}{{\bf{H}}_{\mathcal T,K}}\left( {{{\bf{\Sigma }}_{\mathcal T,1}} +  \cdots  +
{{\bf{\Sigma }}_{\mathcal T,K}}} \right){\bf{H}}_{\mathcal T,K}^H} \right|}}{{\left| {{\bf{I}} + \frac{1}{\sigma^2}{{\bf{H}}_{\mathcal
T,K}}\left( {{{\bf{\Sigma }}_{\mathcal T,1}} + \cdots  + {{\bf{\Sigma }}_{\mathcal T,K - 1}}} \right){\bf{H}}_{\mathcal T,K}^H} \right|}}.
\end{align}

Regarding the power model, as BSs take the main power consumption in the cellular networks, the users' consumed power is not considered here. The power
radiated to the environment for signal transmission is only a portion of BS's total power consumption \cite{Arnold}, so the practical transmit independent power including circuit power, signal processing power, cooling loss etc. at the BS should be taken into account. For the BS deployed with multiple antennas, the transmit independent power is mainly related to the number of active transmit antennas, i.e., when an antenna is on, the
corresponding active RF chain consumes circuit, processing power etc... Thus, as a good approximation of the practical power model, we consider a general model given by
\begin{equation} \label{eq:PowerModel}
\begin{array}{l}
\displaystyle P_{\rm{total}} = f(P,M_{{a}}),
\end{array}
\end{equation}
which is mainly related to the transmit power and number of active transmit antennas. We can assume that $f(P,M_{\rm{a}})$ is monotonously increasing as a function of $P$
and $M_{\rm{a}}$, respectively and based on \cite{fractionalprogramming}, we also assume that $f(P,M_{\rm{a}})$ is affine or convex as a function
of $P$ and $M_a$. Motivated by \cite{Xu1,Arnold} and more specifically, we consider an affine power consumption model, which can be denoted
as
\begin{equation} \label{eq30}
\begin{array}{l}
\displaystyle P_{\rm{total}} = \frac{P}{\eta} + M_a P_{\rm{dyn}} + P_{\rm{sta}},
\end{array}
\end{equation}
where $\eta$ denotes the PA efficiency; $M_a P_{\rm{dyn}}$ denotes the dynamic power consumption proportional to the number of
active transmit antennas, e.g. circuit power of corresponding RF chains which is always proportional to $M_a$; and $P_{\rm{sta}}$ accounts for the static
power independent of both $M_a$ and $P$ which includes power consumption of the baseband processing, battery unit etc.. $M_a P_{\rm{dyn}} + P_{\rm{sta}}$ in total is the transmit independent power.

Note that although the optimization procedure is performed based on the affine model, the idea can be simply extended to other convex power
model case, e.g. considering the rate dependent $P_{\rm{sta}}$ like \cite{Isheden1}. Meanwhile, note that we omit the effect of the signal processing power running the proposed algorithms in the power model, as it is practical that the
$P_{\rm{sta}}$ and $P_{\rm{dyn}}$ would be significantly larger than the signal processing power of the algorithms.

\subsection{Problem Formulation}

The EE is defined as the sum-rate divided by the total power consumption. The maximum sum-rate capacity of MIMO-BC is achieved by DPC, and thus we consider the
DPC achieving sum-rate capacity in this paper. It is worthwhile noting that there exists a performance gap between the capacity and the actual rate achieved by the
cellular networks due to practical constraints such as acquiring of CSIT, overhead of pilots, and practical coding and modulation schemes{\footnote{The acquiring of CSIT, overhead of pilots would have non-trivial effects on the EE performance, because the overhead and imperfect CSIT not only causes capacity decrease but also induces extra energy consumption. The effect of these practical considerations is left for the future work.}}. Nevertheless, DPC achieving sum-rate capacity is the performance upper bound for the MIMO-BC, which helps to reveal the theoretical limits, and thus is employed in this paper.

The objective of this paper is to maximize the EE of MIMO-BC. Based on the sum-rate (\ref{eq:sysmodel:1}) and the total power consumption model (\ref{eq30}) and noting that $P = \sum \nolimits_{i=1}^{K} {\rm Tr}\left({{\bf{\Sigma }}_{\mathcal T,i}}\right)$ and $M_a = |{\cal T}|$, the optimization problem can be defined as
\begin{align}
\max \limits_{{\cal T}, {{\bf{\Sigma }}_{\mathcal T,1}},\ldots,{{\bf{\Sigma }}_{\mathcal T,K}}:{{\bf{\Sigma }}_{\mathcal T,i}} \succeq 0, i =1,\ldots,K} ~& \frac{C_{{\rm{BC}}}\left( {{\bf{H}}_{\mathcal T,1}},\ldots,{{\bf{H}}_{\mathcal T,K}}, {{\bf{\Sigma }}_{\mathcal T,1}},\ldots,{{\bf{\Sigma }}_{\mathcal T,K}}  \right)}{\frac{\sum \nolimits_{i=1}^{K} {\rm Tr}\left({{\bf{\Sigma }}_{\mathcal T,i}}\right)}{\eta } +
|{\cal T}|{P_{{\rm{dyn}}}} + {P_{{\rm{sta}}}}}, \label{eq2:problem} \\
{\rm subject~to}~& 
\sum \nolimits_{i=1}^{K} {\rm Tr}\left({{\bf{\Sigma }}_{\mathcal T,i}}\right) \leq P_{\max},\label{eq2:problem:c2} \\
~&C_{{\rm{BC}}}\left( {{\bf{H}}_{\mathcal T,1}},\ldots,{{\bf{H}}_{\mathcal T,K}}, {{\bf{\Sigma }}_{\mathcal T,1}},\ldots,{{\bf{\Sigma }}_{\mathcal T,K}}  \right) \geq {C_{{\rm{min}}}},\label{eq2:problem:c3}
\end{align}
where $P_{\max}$ and $C_{\min}$ are the maximum sum transmit power constraint at the BS and
minimum sum-rate constraint, respectively. However, the solution of the above problem
is nontrivial since the objective function is nonconcave even under fixed ${\cal T}$. Fortunately,
we can utilize the duality between MIMO-BC and MIMO-MAC to simplify the problem formulation.

The transmission model of the dual MIMO-MAC can be denoted as
\begin{equation} \label{eq2}
\begin{array}{l}
{\bf{y}}_{\rm{MAC}} = \sum \limits _{i=1}^{K}{\bf{H}}_{{\mathcal T},i}^{H} {\bf{u}}_i + {\bf{n}},
\end{array}
\end{equation}
where ${\bf{n}} \in {\mathbb{C}}^{M_a \times 1}$ is the independent Gaussian noise with each entry ${\mathcal{CN}}(0,\sigma^2)$, ${\bf{u}}_i \in {\mathbb C}^{N \times 1}$ is the transmitted signal of user $i$ and ${{\bf{Q}}_{\mathcal T,i}} = {\mathbb E}\left( {\bf{u}}_i {\bf{u}}_i^H\right)$ is the transmit covariance matrix of user $i$. Given ${{\bf{Q}}_{\mathcal T,i}},i=1,\ldots,K$, the sum-rate of the MIMO-MAC can be denoted as
\begin{align} \label{eq5}
 {C_{{\rm{MAC}}}}\left( {{{\bf{H}}_{\mathcal T,1}^H}, \ldots ,{{\bf{H}}_{\mathcal T,K}^H},{{\bf{Q}}_{\mathcal T,1}},\ldots,{{\bf{Q}}_{\mathcal T,K}}  } \right)
 =  W\log \left| {{\bf{I}} + \frac{1}{\sigma^2}\sum\limits_{i = 1}^K
{{\bf{H}}_{\mathcal T,i}^H{{\bf{Q}}_{\mathcal T,i}}{{\bf{H}}_{\mathcal T,i}}} } \right|.
\end{align}

According to the duality between MIMO-BC and MIMO-MAC \cite{Duality}, we have that for any MIMO-BC (\ref{eq1}) with transmit covariance ${{\bf{\Sigma }}_{\mathcal T,i}}, i=1,\ldots,K$, there exists a dual MIMO-MAC (\ref{eq2}) with transmit covariance ${{\bf{Q}}_{\mathcal T,1}},\ldots,{{\bf{Q}}_{\mathcal T,K}}$ using the same sum transmit power
\begin{align} \label{eq:dual1}
\sum \nolimits_{i=1}^{K} {\rm Tr}\left({{\bf{\Sigma }}_{\mathcal T,i}}\right) = \sum \nolimits_{i=1}^{K} {\rm Tr}\left({{\bf{Q }}_{\mathcal T,i}}\right)
\end{align}
such that
\begin{align} \label{eq:dual2}
 {C_{{\rm{MAC}}}}\left( {{{\bf{H}}_{\mathcal T,1}^H}, \ldots ,{{\bf{H}}_{\mathcal T,K}^H},{{\bf{Q}}_{\mathcal T,1}},\ldots,{{\bf{Q}}_{\mathcal T,K}}  } \right)
=
 {C_{{\rm{BC}}}}\left( {{{\bf{H}}_{\mathcal T,1}}, \ldots ,{{\bf{H}}_{\mathcal T,K}},{{\bf{\Sigma}}_{\mathcal T,1}},\ldots,{{\bf{\Sigma}}_{\mathcal T,K}}  } \right),
\end{align}
and {\it vice versa}. Therefore, based on (\ref{eq5})(\ref{eq:dual1}) and (\ref{eq:dual2}), the problem (\ref{eq2:problem}) is reformulated as
\begin{align}
\max \limits_{{\cal T}, {{\bf{Q }}_{\mathcal T,1}},\ldots,{{\bf{Q }}_{\mathcal T,K}}:{{\bf{Q }}_{\mathcal T,i}} \succeq 0, i =1,\ldots,K} ~& \frac{W\log \left| {{\bf{I}} + \frac{1}{\sigma^2}\sum\limits_{i = 1}^K
{{\bf{H}}_{\mathcal T,i}^H{{\bf{Q}}_{\mathcal T,i}}{{\bf{H}}_{\mathcal T,i}}} } \right|}{\frac{\sum \nolimits_{i=1}^{K} {\rm Tr}\left({{\bf{Q }}_{\mathcal T,i}}\right)}{\eta } +
|{\cal T}|{P_{{\rm{dyn}}}} + {P_{{\rm{sta}}}}}, \label{eq2:dualproblem} \\
{\rm subject~to}~& 
\sum \nolimits_{i=1}^{K} {\rm Tr}\left({{\bf{Q }}_{\mathcal T,i}}\right) \leq P_{\max}\label{eq2:dualproblem:c2} \\
~&W\log \left| {{\bf{I}} + \frac{1}{\sigma^2}\sum\limits_{i = 1}^K
{{\bf{H}}_{\mathcal T,i}^H{{\bf{Q}}_{\mathcal T,i}}{{\bf{H}}_{\mathcal T,i}}} } \right| \geq {C_{{\rm{min}}}}.\label{eq2:dualproblem:c3}
\end{align}

If we can obtain the optimal solution ${\cal T}_{\rm opt}, {{\bf{Q }}^{\rm opt}_{\mathcal T,1}},\ldots,{{\bf{Q }}^{\rm opt}_{\mathcal T,K}}$ of (\ref{eq2:dualproblem}), the optimal ${\cal T}_{\rm opt}, {{\bf{\Sigma }}^{\rm opt}_{\mathcal T,1}},\ldots,{{\bf{\Sigma }}^{\rm opt}_{\mathcal T,K}}$ for problem (\ref{eq2:problem}) can be corresponding determined based on the mapping in \cite[Sec. IV-B]{Duality}. Therefore, we will focus on optimizing (\ref{eq2:dualproblem}) in the rest of this paper.

Look at problem (\ref{eq2:dualproblem}) then. Since ${\cal T}$ affects the EE in a comprehensive manner, i.e., ${\cal T}$ is related to both channel matrices and
the dynamic power consumption, solving ${\cal T}$ jointly with ${{\bf{Q }}_{\mathcal T,1}},\ldots,{{\bf{Q }}_{\mathcal T,K}}$ is not straightforward. Furthermore, under fixed ${\cal T}$, the optimization problem (\ref{eq2:dualproblem}) becomes a concave fractional program (also quasi-concave), for which convex optimization techniques are applicable. Since for any optimization problems, we can first optimize over some of the variables and then over the remaining ones \cite[Sec. 4.1.3, p. 133]{Boyed}, we will optimize the transmit covariances at first under fixed ${\cal T}$ and then employ ATAS technique to determine ${\cal T}$. In the next two sections, we will discuss these two techniques respectively.


\section{EE Optimization Under Fixed Transmit Antenna Set}\label{Sec3}

In this section, we will derive the optimal energy efficient transmit covariances under
fixed $\cal T$. Let us look at (\ref{eq2:dualproblem}) with fixed $\mathcal T$ again.
Since the numerator is concave and the denominator is affine, (\ref{eq2:dualproblem}) is a
quasiconcave optimization problem, which can be solved through the standard convex optimization
techniques, i.e., interior-point methods \cite{Boyed}. However, the numerical methods would
be still too complex when the user number becomes significantly large. Thus, developing well
structured algorithms is necessary. For ease
of description, we omit $\cal T$ in the subscript in this section and rewrite the optimization
problem as
\begin{align}
\max \limits_{{{\bf{Q }}_{1}},\ldots,{{\bf{Q }}_{K}}:{{\bf{Q }}_{i}} \succeq 0, i =1,\ldots,K} ~& \frac{W\log \left| {{\bf{I}} + \frac{1}{\sigma^2}\sum\limits_{i = 1}^K
{{\bf{H}}_{i}^H{{\bf{Q}}_{i}}{{\bf{H}}_{i}}} } \right|}{\frac{\sum \nolimits_{i=1}^{K} {\rm Tr}\left({{\bf{Q }}_{i}}\right)}{\eta } +
M_a{P_{{\rm{dyn}}}} + {P_{{\rm{sta}}}}}, \label{eq2:dualproblem:sec3} \\
{\rm subject~to}~& 
\sum \nolimits_{i=1}^{K} {\rm Tr}\left({{\bf{Q}}_{i}}\right) \leq P_{\max},\label{eq2:dualproblem:c2:sec3} \\
~&W\log \left| {{\bf{I}} + \frac{1}{\sigma^2}\sum\limits_{i = 1}^K
{{\bf{H}}_{i}^H{{\bf{Q}}_{i}}{{\bf{H}}_{i}}} } \right| \geq {C_{{\rm{min}}}}.\label{eq2:dualproblem:c3:sec3}
\end{align}

Before solving the problem (\ref{eq2:dualproblem:sec3}), we introduce an auxiliary function
\begin{align}
f(P) = \max \limits_{{{\bf{Q }}_{1}},\ldots,{{\bf{Q }}_{K}}:{{\bf{Q }}_{i}} \succeq 0, i =1,\ldots,K, \sum \nolimits_{i=1}^{K} {\rm Tr}\left({{\bf{Q }}_{i}}\right) = P}  W\log \left| {{\bf{I}} + \frac{1}{\sigma^2}\sum\limits_{i = 1}^K
{{\bf{H}}_{i}^H{{\bf{Q}}_{i}}{{\bf{H}}_{i}}} } \right|, \label{eq:auxiliary}
\end{align}
where $P$ is the sum transmit power. Since $W\log \left| {{\bf{I}} + \frac{1}{\sigma^2}\sum\limits_{i = 1}^K
{{\bf{H}}_{i}^H{{\bf{Q}}_{i}}{{\bf{H}}_{i}}} } \right|$ is concave, $f(P)$ is nondecreasing and concave as a
function of $P$ according to \cite[Lemma 5]{Chong3}. Furthermore, the objective function of (\ref{eq2:dualproblem:sec3}) is equivalent to
$\xi(P)=\frac{f(P)}{\frac{P}{\eta } + M_a{P_{{\rm{dyn}}}} + {P_{{\rm{sta}}}}}$, which is a quasi-concave
function, and we can simply verify that there exists a globally optimal point $P^*$ to maximize  $\xi(P)$, and $\xi(P)$ is monotonously nondecreasing
when $P<P^*$ and monotonously nonincreasing when $P>P^*$.

Based on the above feature, solving (\ref{eq2:dualproblem:sec3}) can be transformed into solving the following three subproblems, where {\bf P1}
is the unconstrained EE optimization problem, {\bf P2} is the sum-rate maximization problem under sum transmit power constraint which relates to the constraint (\ref{eq2:dualproblem:c2:sec3}), and {\bf P3}  is the sum transmit power  minimization problem under sum-rate constraint which relates to the constraint (\ref{eq2:dualproblem:c3:sec3}).
\begin{align}
{\bf P1:}~\{{\bf{Q}}_i^*\}_{i=1}^K = \arg \max \limits_{{{\bf{Q }}_{1}},\ldots,{{\bf{Q }}_{K}}:{{\bf{Q }}_{i}} \succeq 0, i =1,\ldots,K} ~& \frac{W\log \left| {{\bf{I}} + \frac{1}{\sigma^2}\sum\limits_{i = 1}^K
{{\bf{H}}_{i}^H{{\bf{Q}}_{i}}{{\bf{H}}_{i}}} } \right|}{\frac{\sum \nolimits_{i=1}^{K} {\rm Tr}\left({{\bf{Q }}_{i}}\right)}{\eta } +
M_a{P_{{\rm{dyn}}}} + {P_{{\rm{sta}}}}} \label{eq2:P1}
\end{align}
\begin{align}
{\bf P2:}~\{\hat{\bf{Q}}_i\}_{i=1}^K =  ~& \arg \max \limits_{{{\bf{Q }}_{1}},\ldots,{{\bf{Q }}_{K}}:{{\bf{Q }}_{i}} \succeq 0, i =1,\ldots,K, \sum \nolimits_{i=1}^{K} {\rm Tr}\left({{\bf{Q}}_{i}}\right) = P_{\max}} \frac{W\log \left| {{\bf{I}} + \frac{1}{\sigma^2}\sum\limits_{i = 1}^K
{{\bf{H}}_{i}^H{{\bf{Q}}_{i}}{{\bf{H}}_{i}}} } \right|}{\frac{\sum \nolimits_{i=1}^{K} {\rm Tr}\left({{\bf{Q }}_{i}}\right)}{\eta } +
M_a{P_{{\rm{dyn}}}} + {P_{{\rm{sta}}}}} \nonumber \\
= ~& \arg \max \limits_{{{\bf{Q }}_{1}},\ldots,{{\bf{Q }}_{K}}:{{\bf{Q }}_{i}} \succeq 0, i =1,\ldots,K, \sum \nolimits_{i=1}^{K} {\rm Tr}\left({{\bf{Q}}_{i}}\right) \leq P_{\max}}{W\log \left| {{\bf{I}} + \frac{1}{\sigma^2}\sum\limits_{i = 1}^K
{{\bf{H}}_{i}^H{{\bf{Q}}_{i}}{{\bf{H}}_{i}}} } \right|} \label{eq2:P2}
\end{align}
\begin{align}
{\bf P3:}~\{\bar{\bf{Q}}_i\}_{i=1}^K =  ~& \arg \max \limits_{{{\bf{Q }}_{1}},\ldots,{{\bf{Q }}_{K}}:{{\bf{Q }}_{i}} \succeq 0, i =1,\ldots,K, W\log \left| {{\bf{I}} + \frac{1}{\sigma^2}\sum\limits_{i = 1}^K
{{\bf{H}}_{i}^H{{\bf{Q}}_{i}}{{\bf{H}}_{i}}} } \right| = {C_{{\rm{min}}}}} \frac{W\log \left| {{\bf{I}} + \frac{1}{\sigma^2}\sum\limits_{i = 1}^K
{{\bf{H}}_{i}^H{{\bf{Q}}_{i}}{{\bf{H}}_{i}}} } \right|}{\frac{\sum \nolimits_{i=1}^{K} {\rm Tr}\left({{\bf{Q }}_{i}}\right)}{\eta } +
M_a{P_{{\rm{dyn}}}} + {P_{{\rm{sta}}}}} \nonumber \\
=  ~& \arg \min \limits_{{{\bf{Q }}_{1}},\ldots,{{\bf{Q }}_{K}}:{{\bf{Q }}_{i}} \succeq 0, i =1,\ldots,K, W\log \left| {{\bf{I}} + \frac{1}{\sigma^2}\sum\limits_{i = 1}^K
{{\bf{H}}_{i}^H{{\bf{Q}}_{i}}{{\bf{H}}_{i}}} } \right| \geq {C_{{\rm{min}}}}}\sum \limits_{i=1}^{K} {\rm Tr}\left({{\bf{Q }}_{i}}\right) \label{eq2:P3}
\end{align}
Based on {\bf P1} we have that $P^* = \sum\limits_{i =
1}^K {{\rm{Tr}}\left( {{{\bf{Q}}^*_{i}}} \right)}$ is the globally optimal sum transmit power, and based on {\bf P3} we have that $\bar P = \sum\limits_{i = 1}^K {{\rm{Tr}}\left( {{\bar{\bf{Q}}_{i}}} \right)}$ is the minimum sum transmit power fulfilling the sum-rate constraint, while $P_{\max} = \sum\limits_{i = 1}^K {{\rm{Tr}}\left( {{\hat{\bf{Q}}_{i}}} \right)}$ is the maximum sum transmit power for {\bf P2}. Therefore, according to the feature that $\xi(P)$ is nondecreasing when $P<P^*$ and nonincreasing when $P>P^*$, we have the optimal solution for (\ref{eq2:dualproblem:sec3}) given by
\begin{equation} \label{eq102}
\begin{array}{l}{{\bf Q}^{{\rm{opt}}}_i} = \left\{ {\begin{array}{*{20}{l}}
{\bf{Q}}^*_i,\\
\hat {\bf{Q}}_i,\\
\bar {\bf{Q}}_i,\\
{\rm infeasible},
\end{array}} \right.\begin{array}{*{20}{l}}
{\bar P \le {P^*} \le {P_{\max }}}\\
{{P^*} \ge {P_{\max }} \ge \bar P}\\
{{P^*} \le \bar P \le {P_{\max }}}\\
\bar P > {P_{\max }}
\end{array}, i=1,\ldots,K.\end{array}
\end{equation}
The above solution (\ref{eq102}) indicates at first that $\bar P \le {P_{\max }}$ is required to guarantee the feasibility of solution for (\ref{eq2:dualproblem:sec3}), and $[\bar P , {P_{\max }}]$ is the feasible region of the sum transmit power. When ${\bar P \le {P^*} \le {P_{\max }}}$, globally EE optimal point is involved in the feasible region, and thus employing ${\bf{Q}}^*_i,i=1,\ldots,K$ is globally optimal. When ${{P^*} \ge {P_{\max }} \ge \bar P}$, $\xi(P)$ is nondecreasing in $[\bar P, {P_{\max }}]$, and thus ${P_{\max }}$ and corresponding $\hat {\bf{Q}}_i,i=1,\ldots,K$ are optimal. When ${{P^*} \le \bar P \le {P_{\max }}}$, $\xi(P)$ is nonincreasing in $[\bar P, {P_{\max }}]$, and thus $\bar P$ and corresponding $\bar {\bf{Q}}_i,i=1,\ldots,K$ are optimal. Note that when the solution is infeasible, we will choose ${\bf{Q}}^{\rm{opt}}_i = \hat{\bf{Q}}_i, i=1,\ldots,K$. This choice has practical significance, as when the sum-rate constraint is infeasible, the best choice is trying to achieve the maximum sum-rate.

The remaining thing for the optimization is to solve problem {\bf P1}-{\bf P3}. Fortunately, {\bf P2} can be solved by the spectral efficient iterative water-filling schemes efficiently in \cite{IterativeBC}, while {\bf P3} can be solved efficiently in \cite{PowerMin}. Therefore, we only need to address the unconstrained EE optimization problem {\bf P1}, which is presented in the following subsection.

%
%
%

%
%
%
%
%
%
%
%
%
%

\subsection{Unconstrained EE Optimization}\label{sec3aa}


To solve {\bf P1}, a choice is utilizing the idea of nested optimization \cite{Chong3}. Taking (\ref{eq:auxiliary}) into (\ref{eq2:P1}), {\bf P1} can be rewritten as
\begin{align}
\max \limits_{P:P\geq 0}~& \frac{f(P)}{\frac{P}{\eta } + M_a{P_{{\rm{dyn}}}} + {P_{{\rm{sta}}}}}.  \label{eq:equi}
\end{align}
Through employing bisection for (\ref{eq:equi}) jointly with spectral efficient iterative water-filling \cite{IterativeBC} for (\ref{eq:auxiliary}), {\bf P1} can be solved efficiently. However, bisection cannot give insights on the solution structure. Thus, we will propose a more efficient scheme, i.e. energy efficient iterative water-filling.

The energy efficient iterative water-filling is motivated by the spectral efficient scheme in \cite{IterativeBC}, in which block-coordinate ascent
algorithm \cite[Sec. 2.7]{NonlinearProgramming} is utilized. If we can write the EE as the similar structure with the block-coordinate
ascent algorithm and then prove it satisfies the conditions of \cite[Sec. 2.7]{NonlinearProgramming}, we can also obtain an iterative solution for
problem {\bf P1}.



We define the following function $g \left( \cdot \right)$ at first.
\begin{align}
g \left( {{{\bf{Q}}_1}, \ldots ,{{\bf{Q}}_K}} \right) = \frac{W \log \left| {{\bf{I}} + \frac{1}{\sigma^2}\sum\limits_{i = 1}^K
{{\bf{H}}_i^H{{\bf{Q}}_i}{{\bf{H}}_i}} } \right|}{\frac{{\sum\nolimits_{i = 1}^K{{\rm{Tr}}\left( {{{\bf{Q}}_i}} \right)}}}{\eta} + M_a
P_{\rm{dyn}} + P_{\rm{sta}}}\label{eq8}
\end{align}

For the block-coordinate ascent algorithm, given the current iteration ${{\bf{Q}}^{(k)}} = \left({{{\bf{Q}}_1^{(k)}}, \ldots
,{{\bf{Q}}_K^{(k)}}}\right)$, the next iteration ${{\bf{Q}}^{(k+1)}} = \left({{{\bf{Q}}_1^{(k+1)}}, \ldots ,{{\bf{Q}}_K^{(k+1)}}}\right)$ can be
generated as
\begin{equation} \label{eq801}\begin{array}{l}
{{\bf{Q}}_i^{(k+1)}}  =\displaystyle \arg \max \limits_{{{\bf{Q}}_i}:{\bf{Q}}_i \succeq 0} g \left( {{{\bf{Q}}_1^{(k+1)}}, \ldots
,{{\bf{Q}}_{i-1}^{(k+1)}},
 {{\bf{Q}}_i},{{\bf{Q}}_{i+1}^{(k)}},\ldots,{{\bf{Q}}_K^{(k)}}}
 \right).
\end{array}
\end{equation}
To apply the iterative algorithm efficiently, the following two conditions should be satisfied. For one thing, the solution of ({\ref{eq801}})
should be uniquely attained \cite[Proposition 2.7.1]{NonlinearProgramming}. For another, the solution should be simple and easy to implement.
Fortunately, the two conditions are both satisfied and the solution can be obtained following an energy efficient water-filling feature. We are
interested in presenting it as follows.

%
%
%
%
%
%

\subsubsection{Energy Efficient Water-filling} \label{EEW}

Based on \cite{IterativeBC,IterativeMAC}, it is fulfilled that
\begin{align}
& \log \left| {{\bf{I}} +
\frac{1}{\sigma^2}\sum\limits_{i = 1}^K {{\bf{H}}_i^H{{\bf{Q}}_i}{{\bf{H}}_i}} } \right| \nonumber
= \log \left| {{\bf{I}} + \frac{1}{\sigma^2}\sum\limits_{j \ne i} {{\bf{H}}_j^H{{\bf{Q}}_j}{{\bf{H}}_j}} } \right|  \\ &+ \log \left| {\bf{I}} +
{{\left( \sigma^2{{\bf{I}} + \sum\limits_{j \ne i} {{\bf{H}}_j^H{{\bf{Q}}_j}{{\bf{H}}_j}} } \right)}^{ - 1/2}} \right.  \left.\times
{\bf{H}}_i^H{{\bf{Q}}_i}{{\bf{H}}_i}{{\left( \sigma^2{{\bf{I}} + \sum\limits_{j \ne i} {{\bf{H}}_j^H{{\bf{Q}}_j}{{\bf{H}}_j}} } \right)}^{ -
1/2}}\right| \nonumber \\
 = ~& \log \left| {{{\bf{Z}}_i}} \right| + \log \left| {{\bf{I}} + {\bf{G}}_i^H{{\bf{Q}}_i}{{\bf{G}}_i}} \right|, \label{eq9}
\end{align}
where ${{\bf{Z}}_i} = {\bf{I}} + \frac{1}{\sigma^2}\sum\limits_{j \ne i} {{\bf{H}}_j^H{{\bf{Q}}_j}{{\bf{H}}_j}}$ and ${{\bf{G}}_i} =
{{\bf{H}}_i}{\left( \sigma^2{{\bf{I}} + \sum\limits_{j \ne i} {{\bf{H}}_j^H{{\bf{Q}}_j}{{\bf{H}}_j}} } \right)^{ - 1/2}}$. By denoting
$a_i = \frac{ \sum\limits_{j \neq i
 }{{\rm{Tr}}\left( {{{\bf{Q}}_j}} \right)}}{\eta} + M_a P_{\rm{dyn}} +
P_{\rm{sta}},$$
b_i = W \log \left| {{{\bf{Z}}_i}} \right|$,
and substituting (\ref{eq9}) into (\ref{eq8}) we have that
\begin{equation} \label{eq10}\begin{array}{l}
g \left( {{{\bf{Q}}_1}, \ldots ,{{\bf{Q}}_K}} \right) = \displaystyle \frac{b_i + W\log \left| {{\bf{I}} + {\bf{G}}_i^H{{\bf{Q}}_i}{{\bf{G}}_i}}
\right|}{\frac{{\rm{Tr}}\left( {{{\bf{Q}}_i}} \right)}{\eta} + a_i}
\end{array}
\end{equation}

%

Therefore, we can redefine the problem (\ref{eq801}) by removing the iteration number as to
\begin{equation} \label{eq12}
\begin{array}{l}
\mathop {{\rm{maximize}}}\limits_{{{\bf{Q}}_i}:{\bf{Q}}_i \succeq 0} g \left( {{{\bf{Q}}_1}, \ldots
,{{\bf{Q}}_{i-1}},{{\bf{Q}}_i},{{\bf{Q}}_{i+1}},\ldots,{{\bf{Q}}_K}} \right)
= \displaystyle \mathop {{\rm{maximize}}}\limits_{{{\bf{Q}}_i}:{\bf{Q}}_i \succeq 0} \frac{b_i + W\log \left| {{\bf{I}} + {\bf{G}}_i^H{{\bf{Q}}_i}{{\bf{G}}_i}} \right|}{\frac{{\rm{Tr}}\left( {{{\bf{Q}}_i}}
\right)}{\eta} + a_i}
\end{array}
\end{equation}
by treating ${{{\bf{Q}}_1}, \ldots ,{{\bf{Q}}_{i-1}},{{\bf{Q}}_{i+1}},\ldots,{{\bf{Q}}_K}} $ as constant. 

Since the numerator and denominator in (\ref{eq10}) are concave and affine respectively, (\ref{eq12}) is a concave fractional program. To solve such a problem, it is efficient to relate it to a concave program by separating numerator and denominator with the help of parameter $\lambda$ \cite{frac_program}.

Here we assume that $\lambda$ is nonnegative, and define the parametric problem as
\begin{align}
Y(\lambda) = \max \limits _{{{\bf{Q}}_i}:{{\bf{Q}}_i} \succeq 0} b_i + W\log \left| {{\bf{I}} + {\bf{G}}_i^H{{\bf{Q}}_i}{{\bf{G}}_i}} \right| -\lambda  \left( {{\frac{{{\rm{Tr}}\left(
{{{\bf{Q}}_i}} \right)}}{\eta} + a_i}}\right).\label{eq13}
\end{align}
We have the following two properties, whose proof can be referred to \cite{frac_program} and thus is omitted here.
\begin{lemma}\label{lemma:1}
$Y(\lambda)$ is a strictly decreasing, continuous function on $[0,+\infty)$ where $Y(\lambda) \to +\infty(-\infty)$ if $\lambda \to 0(+\infty)$.
\end{lemma}
\begin{lemma}\label{lemma:2}
Let $\lambda^*$ denote the unique zero of $Y(\lambda)$. The optimal solutions of $Y(\lambda^*)$ and (\ref{eq12}) are the same and $\lambda^*$ is the optimal objective value of (\ref{eq12}).
\end{lemma}

Based on Lemma \ref{lemma:2}, we need to optimize (\ref{eq13}) at first under given $\lambda$ to obtain $Y(\lambda)$ and then solve the equation $Y(\lambda) = 0$ to get the unique $\lambda^*$. In the following, we will optimize (\ref{eq13}) under a given $\lambda$ at first.

%
%
%
%

Perform eigenvalue decomposition on ${\bf{G}}_i^H{{\bf{G}}_i}$ as
\begin{equation} \label{eq1301}
\begin{array}{l}
{\bf{G}}_i^H{{\bf{G}}_i} = {\bf{U}}{\bf{D}}_i{\bf{U}}^H,
\end{array}
\end{equation}
where ${\bf{D}}_i \in{\mathbb C}^{M \times M}$ is diagonal with nonnegative entries and
${\bf{U}} \in{\mathbb C}^{M \times M}$ is unitary. We assume that ${\bf{D}}_i$ has $L$ non-zero diagonal entries ($1 \leq L \leq M$), which
means $[{\bf{D}}_i]_{kk} > 0$ for $k = 1,\ldots,L$ and $[{\bf{D}}_i]_{kk} = 0$ for $k = L+1,\ldots,M$. And then we have the following equation based on $\left|{\bf{I+AB}}\right| = \left|{\bf{I+BA}}\right|$ \cite{IterativeBC}:
\begin{equation} \label{eq16}
\begin{array}{l}
\log \left| {{\bf{I}} + {\bf{G}}_i^H{{\bf{Q}}_i}{{\bf{G}}_i}} \right| = \log \left| {{\bf{I}} + {{\bf{Q}}_i}{\bf{G}}_i^H{{\bf{G}}_i}} \right|
= \log \left| {{\bf{I}} + {{\bf{Q}}_i}{\bf{U}}{\bf{D}}_i{\bf{U}}^H} \right| = \log \left| {{\bf{I}} + {\bf{U}}^H{{\bf{Q}}_i}{\bf{U}}{\bf{D}}_i}
\right|
\end{array}
\end{equation}
Define ${\bf{S}}_i = {\bf{U}}^H{{\bf{Q}}_i}{\bf{U}}$. As ${\bf{U}}$ is unitary, we have that ${\rm{Tr}}({\bf{S}}_i) = {\rm{Tr}}({\bf{Q}}_i)$.
As each ${\bf{S}}_i$ corresponds to a ${\bf{Q}}_i$ via the invertible mapping ${\bf{S}}_i = {\bf{U}}^H{{\bf{Q}}_i}{\bf{U}}$, (\ref{eq13})
is equivalent to solving the following convex optimization problem.
\begin{equation} \label{eq18}
\begin{array}{l}
F(\lambda) = \max \limits _{{{\bf{S}}_i}:{{\bf{S}}_i} \succeq 0} { b_i + W\log \left| {{\bf{I}} + {\bf{S}}_i{{\bf{D}}}_i} \right|} -\lambda \left({{\displaystyle
 \frac{{{\rm{Tr}}\left( {{{\bf{S}}}_i} \right)} }{\eta} + a_i }}\right)
\end{array}
\end{equation}

%
%

It is proved in \cite[Appendix II]{IterativeBC} that the optimal ${{\bf{S}}_i^*}$ for (\ref{eq18}) is diagonal with diagonal elements  $[{\bf{S}}_i^*]_{kk} > 0$
for $k = 1,\ldots,L$ and $[{\bf{S}}_i^*]_{kk} = 0$ for $k = L+1,\ldots,M$. Thus, $F(\lambda)$ with diagonal ${\bf{S}}_i$ is then
\begin{equation} \label{eq19}
\begin{array}{l}
F(\lambda) = \max \limits _{{{\bf{S}}_i}:{{\bf{S}}_i} \succeq 0} b_i + W \sum \limits _{k=1}^{L} \log \left( {{1}} + [{\bf{S}}_i]_{kk}[{{\bf{D}}}_i]_{kk} \right)  -\lambda
\left({{
 \frac{\sum \limits _{k=1}^ L{ {[{{\bf{S}}}_i]_{kk}} } }{\eta} + a_i }}\right).
\end{array}
\end{equation}
As the objective function of (\ref{eq19}) is concave in ${{\bf{S}}_i}$, problem (\ref{eq19}) can be solved by solving the KKT
optimality conditions, and the solution can be denoted as
\begin{align}
[{{\bf{S}}_{i}^*}]_{kk}^{\lambda} = \left[\frac{{\eta  }}{{\ln(2)\lambda}} - \frac{1}{{[{{\bf{D}}_{i}}]_{kk}}}\right]^+,  k = 1,\ldots,L, \label{eq20}
\end{align}
where $[x]^+ = \max(x,0)$. Therefore, under given $\lambda$, the optimal solution for (\ref{eq13}) can be given by
\begin{align}
{{\bf{Q}}^{\lambda}_i} = {\bf{U}}{{\bf{S}}_i^*}^{\lambda }{\bf{U}}^H.
\end{align}

Next, we need to find the unique $\lambda^*$ fulfilling $Y(\lambda^*) = 0$. Since $Y(\lambda)$ and $F(\lambda)$ achieve the same value for any $\lambda$, we can obtain $\lambda^*$
by setting $F(\lambda^*) = 0$, i.e.,
\begin{align} \label{eq21}
{ b_i + \sum \limits _{k=1}^{L} \log \left( {{1}} + \left[\frac{{\eta  }}{{\ln(2)\lambda^*}} -
\frac{1}{{[{{\bf{D}}_{i}}]_{kk}}}\right]^+[{{\bf{D}}}_i]_{kk} \right)}
-\lambda^* \times \left(
 \frac{\sum \limits _{k=1}^ L{ {\left[\frac{{\eta  }}{{\ln(2)\lambda^*}} - \frac{1}{{[{{\bf{D}}_{i}}]_{kk}}}\right]^+} } }{\eta} + a_i \right) = 0.
\end{align}

After obtaining $\lambda^*$ by solving (\ref{eq21}), the optimal solution of (\ref{eq12}) can be derived as
\begin{equation} \label{eq22}
\begin{array}{l}{{\bf{Q}}_i^*} = {\bf{U}}{{\bf{S}}_i^*}^{\lambda ^*}{\bf{U}}^H .
\end{array}
\end{equation}

It is worth noting that the solution in ({\ref{eq20}}) has a water-filling structure where $\lambda$ is the water level, and thus the solution here is referred to as {\it energy efficient water-filling}. Meanwhile, we can employ bisection or Newton's method to determine the water level $\lambda^*$ based on (\ref{eq21}){\footnote{In contrast of solving (\ref{eq21}) directly, an iterative Dinkelbach method \cite{fractionalprogramming} can also be applied to obtain $\lambda^*$.}}. Moreover, since $\lambda^*$ is unique based on Lemma \ref{lemma:1}, the solution of (\ref{eq12}) is unique and globally optimal. Thus, block-coordinate ascent algorithms can be efficiently applied.

%



\subsubsection{Iterative Algorithm}

Based on the derivation in section \ref{EEW} and the block-coordinate ascent algorithm, the energy efficient iterative water-filling scheme can
be denoted as follows.
\begin{algorithm}\label{alg:1}
\textbf{Energy Efficient Iterative Water-filling}
\begin{itemize}
    \item \textbf{Initialization}: Set $j=1$, $\xi_0 = 0$ and  ${{\bf{Q}}_i} = {\bf{0}}, i = 1,\ldots,K.$, and set the required accuracy as $\Delta$.
    \item \textbf{Repeat}:
    \begin{itemize}
        \item For $i=1:K$

        \begin{enumerate}
            \item Calculate ${{\bf{Q}}_i^*}$ based on the energy efficient water-filling algorithm (\ref{eq20})(\ref{eq21}) and (\ref{eq22})
            \item Update ${\bf{Q}}_i$ as ${{\bf{Q}}_i^*}$
        \end{enumerate}
        \item End
        \item Calculate the EE $\xi_{j}$ and $j++$
    \end{itemize}
    \item \textbf{Until $\xi_{j}-\xi_{j-1} \leq \Delta$.}
\end{itemize}
\end{algorithm}

The {\it proof of convergence} is given as follows.

\begin{proof}
Firstly, during each step, the energy efficient water-filling can achieve global maximum treating the other
users' transmit covariance matrices as constant, so the EE is non-decreasing within each step. As the EE is bounded, the EE converges to a limit.
Secondly, since the derivation of each step is unique, the set of ${\bf{Q}}_1,\ldots,{\bf{Q}}_K$ also converges to a limit based on \cite[Sec.
2.7]{NonlinearProgramming}. Moreover, {\bf P1} is a concave fractional program, and for the concave fractional programs, any local maximum is a global maximum \cite{frac_program}. Thus, the energy efficient
iterative water-filling converges to the global optimality for {\bf P1}.
\end{proof}

\begin{figure}[t]
\begin{center}
\includegraphics[width=12.5cm] {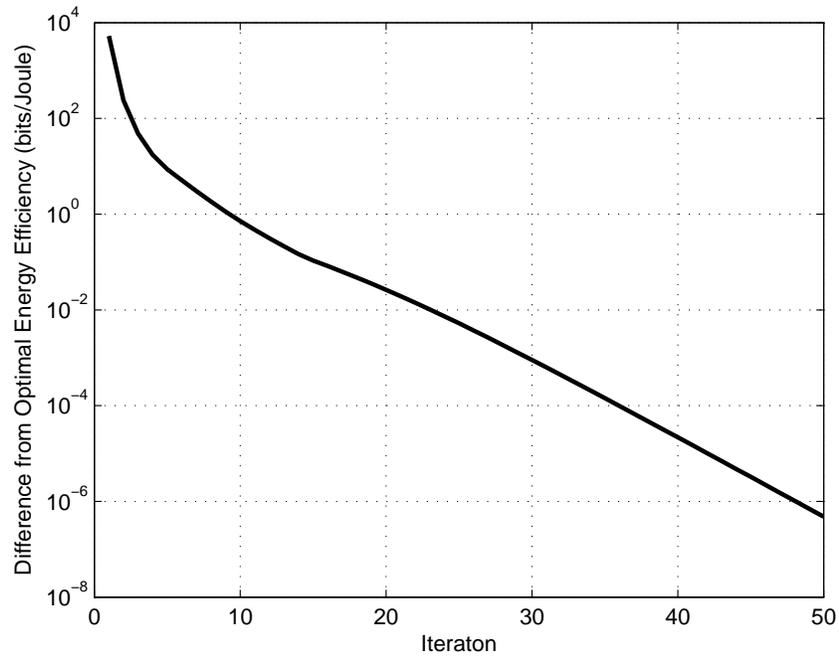}
\end{center}
\caption{EE convergence behavior of the proposed scheme.} \label{fig001}
\end{figure}

Note that as the proof does not depend on the starting point, we can start the algorithm from any values of
${\bf{Q}}_1,\ldots,{\bf{Q}}_K$. To show the efficiency of the proposed scheme, we give the simulation results in Fig. \ref{fig001}. The convergence
behavior of the proposed scheme is shown with setting $d=1$km, $M=4$, $N=4$, $K=10$, and the channels are generated randomly 10 times. The achieved maximum EE is about 4.422$\times 10^5$bits/Joule.  We can see that our proposed iterative scheme converges almost exponentially fast.

\subsection{Complexity Analysis}\label{sec3complex}

In this subsection, we provide complexity analysis for the proposed transmit covariance optimization algorithm. As the algorithm separates the optimizations into three subproblems, i.e. the unconstrained EE optimization problem (\ref{eq2:P1}), sum-rate maximization problem (\ref{eq2:P2}) and power minimization problem (\ref{eq2:P3}), we will provide the complexity of each subproblem at first. The complexity of the sum-rate maximization (\ref{eq2:P2}) via spectral efficient iterative water-filling is evaluated in \cite{IterativeBC}, which increases {\it linearly} with $K$, the number of users, and is a extremely desirable property when considering systems are with large number of users. For the optimization of (\ref{eq2:P3}), a bisection procedure is applied in \cite{PowerMin} which employs the sum-rate maximization iterative water-filling in each iteration. The algorithm in \cite{PowerMin} also has a linear complexity with $K$, although the run time would be higher than the iterative water-filling for (\ref{eq2:P2}) in \cite{IterativeBC}.
About the complexity of the energy efficient iterative water-filling for the unconstrained EE optimization problem (\ref{eq2:P1}), we can conclude that its complexity is also linear with the user number $K$. It is clear that in each iteration, (\ref{eq1301})(\ref{eq20}) and (\ref{eq22}) only require a finite number of subtractions and additions. About the complexity of (\ref{eq21}), a bisection  can be applied to determine the water level $\lambda^*$ here, which can also be performed in linear time with $K$. Moreover, applying bisection methods to deriving $\lambda^*$ in (\ref{eq21}) is similar with deriving the water level in each iteration of the spectral efficient iterative water-filling in \cite{IterativeBC}. Therefore, the proposed energy efficient iterative water-filling has a similar complexity as the spectral efficient iterative water-filling in \cite{IterativeBC}. It is worth noting that employing nested optimization to solve {\bf P1}, i.e., utilizing bisection to solve (\ref{eq:equi}) jointly with spectral efficient iterative water-filling, also has a complexity linear with $K$, but it is more complex than energy efficient iterative water-filling, since in (\ref{eq:equi}) spectral efficient iterative water-filling \cite{IterativeBC} should be applied iteratively for multiple times  due to bisection.

Combining the complexity of three subproblems, the complexity of the proposed transmit covariance optimization is in general linear with user number $K$, which is a desirable property. Another choice to solve the quasi-concave EE optimization is employing the standard convex optimization technique, i.e. standard interior point methods. However, this standard scheme has a complexity that is cubic with respect to the dimensionality of input space (i.e. with respect to $K$, the user number), due to the complexity of inner Newton iterations \cite{IterativeBC,Boyed}. Therefore, compared with the standard interior point methods, our scheme decreases the
complexity significantly.


\section{Active Transmit Antenna Selection}\label{Sec4}

This section will optimize $\mathcal T$ based on the ATAS. The optimization problem is formulated as
\begin{align}
\max \limits_{{\mathcal T} }  \zeta({\mathcal T}),
\end{align}
where
\begin{align}
\zeta({\mathcal T}) = \max \limits_{{{\bf{Q }}_{\mathcal T,1}},\ldots,{{\bf{Q }}_{\mathcal T,K}}:{{\bf{Q }}_{\mathcal T,i}} \succeq 0, i =1,\ldots,K} ~& \frac{W\log \left| {{\bf{I}} + \frac{1}{\sigma^2}\sum\limits_{i = 1}^K
{{\bf{H}}_{\mathcal T,i}^H{{\bf{Q}}_{\mathcal T,i}}{{\bf{H}}_{\mathcal T,i}}} } \right|}{\frac{\sum \nolimits_{i=1}^{K} {\rm Tr}\left({{\bf{Q }}_{\mathcal T,i}}\right)}{\eta } +
|{\cal T}|{P_{{\rm{dyn}}}} + {P_{{\rm{sta}}}}}, \nonumber \\
{\rm subject~to}~& 
\sum \nolimits_{i=1}^{K} {\rm Tr}\left({{\bf{Q }}_{\mathcal T,i}}\right) \leq P_{\max},\nonumber\\
~&W\log \left| {{\bf{I}} + \frac{1}{\sigma^2}\sum\limits_{i = 1}^K
{{\bf{H}}_{\mathcal T,i}^H{{\bf{Q}}_{\mathcal T,i}}{{\bf{H}}_{\mathcal T,i}}} } \right| \geq {C_{{\rm{min}}}}.\label{eq:ATAS}
\end{align}
From the standpoint of SE, activating all transmit antennas is always optimal. However, this conclusion does not hold under the energy efficient scenario. As more active transmit antennas achieve higher sum-rate at the cost of higher dynamic power, there exists a tradeoff between the power consumption cost and the sum-rate gain. Thus, ATAS is necessary. The ATAS here is different from the spectral efficient transmit antenna selection, as the conventional transmit antenna selection is always utilized in the scenario when the number of transmit antennas is larger than the RF chains and the purpose is to employ the selection diversity. Although the ATAS here can also acquire the selection diversity, its main
purpose is to choose appropriate active transmit antenna set to exploit the best tradeoff between the dynamic power consumption cost and the sum-rate gain.
After determining the active transmit antenna set, inactive antennas should be turned off through micro-sleep or DTX.

It is intuitive to see that the exhaustive search  is the optimal ATAS scheme. For each possible ${\mathcal T} \subseteq \{1,\ldots,M\}$, the BS calculates the EE based on the algorithm in section \ref{Sec3}, and then chooses the optimal active transmit antenna set as follows after comparing the EE.
\begin{align}
{\mathcal T}_{\rm{opt}} = {\rm{arg}} \mathop {\max} \limits_{{\mathcal T} \subseteq \{1,\ldots,M\}}  \zeta({\mathcal T})
\end{align}
During the comparison, if (\ref{eq:ATAS}) is infeasible, $\zeta({\mathcal T})$ is then set as zero. If there is no feasible active transmit antenna set, we would choose ${\mathcal T}_{\rm{opt}} = \{1,\ldots,M\}$ to achieve the highest sum-rate. Nevertheless, the complexity of the exhaustive search ATAS is
too high to implement. Thus, developing schemes with low complexity is of importance.

Let us look at the problem formulation again. Interestingly, given a constant $M_a =|{\mathcal T}|$, we can have the following approximation for EE at first.
\begin{align}
\mathop {\max }\limits_{{\mathcal T}:|{\mathcal T}|=M_a}\mathop {\max }\limits_{{\bf Q}} \frac{W \log \left|
{{\bf{I}} + \frac{1}{\sigma^2}\sum\limits_{i = 1}^K {{\bf{H}}_{\mathcal T,i}^H{{\bf{Q}}_{\mathcal T,i}}{{\bf{H}}_{\mathcal T,i}}} }
\right|}{\frac{{\sum\nolimits_{i = 1}^K{{\rm{Tr}}\left( {{{\bf{Q}}_{\mathcal T,i}}} \right)}}}{\eta} + |\mathcal T| P_{\rm{dyn}} +
P_{\rm{sta}}} \nonumber
~&\mathop \ge \limits^{(a)}
  \mathop {\max }\limits_{{\mathcal T}:|{\mathcal T}|=M_a}\mathop {\max }\limits_{{P}} \frac{W \log \left|
{{\bf{I}} + \frac{P}{NK\sigma^2}\sum\limits_{i = 1}^K {{\bf{H}}_{\mathcal T,i}^H{{\bf{H}}_{\mathcal T,i}}} }
\right|}{\frac{P}{\eta} + |\mathcal T| P_{\rm{dyn}} +
P_{\rm{sta}}} \nonumber
\\
&\mathop \ge \limits^{(b)}
  \mathop {\max }\limits_{{\mathcal T}:|{\mathcal T}|=M_a}\mathop {\max }\limits_{{P}} \frac{W \log \left|
{\frac{P}{NK\sigma^2}\sum\limits_{i = 1}^K {{\bf{H}}_{\mathcal T,i}^H{{\bf{H}}_{\mathcal T,i}}} }
\right|}{\frac{P}{\eta} + |\mathcal T| P_{\rm{dyn}} +
P_{\rm{sta}}}\label{eq:ATAS:1}
\end{align}
where (a) assumes equal transmit power allocation at each antenna and (b) assumes the systems act in high signal to noise ratio (SNR) regime. And (\ref{eq:ATAS:1}) is equivalent to optimizing
\begin{align}
\mathop {\max }\limits_{{\mathcal T}:|{\mathcal T}|=M_a} \left|
{\sum\limits_{i = 1}^K {{\bf{H}}_{\mathcal T,i}^H{{\bf{H}}_{\mathcal T,i}}} }
\right| =   \mathop {\max }\limits_{{\mathcal T}:|{\mathcal T}|=M_a} \left|
{ {{\bf{H}}_{\mathcal T}^H{{\bf{H}}_{\mathcal T}}} }
\right|.
\end{align}
Similarly, we can have similar approximation for sum-rate, and then optimizing
\begin{align}
\mathop {\max }\limits_{{\mathcal T}:|{\mathcal T}|=M_a} W \log \left|
{{\bf{I}} + \frac{1}{\sigma^2}\sum\limits_{i = 1}^K {{\bf{H}}_{\mathcal T,i}^H{{\bf{Q}}_{\mathcal T,i}}{{\bf{H}}_{\mathcal T,i}}} }
\right|
\end{align}
is equivalent to optimizing
\begin{align}
   \mathop {\max }\limits_{{\mathcal T}:|{\mathcal T}|=M_a} \left|
{ {{\bf{H}}_{\mathcal T}^H{{\bf{H}}_{\mathcal T}}} }
\right|
\end{align}


Thus, we can choose the active transmit antenna set based on the criterion of $\left|
{ {{\bf{H}}_{\mathcal T}^H{{\bf{H}}_{\mathcal T}}} }
\right|$ under given $M_a$. However, calculating $\left|
{ {{\bf{H}}_{\mathcal T}^H{{\bf{H}}_{\mathcal T}}} }
\right|$ under given $M_a$ still requires calculation of matrix determinant. Motivated by \cite{RXSelBD}, although the channel Frobenius norm cannot characterize the
determinant completely, it is related to the determinant because the Frobenius norm indicates the overall energy of the channel, i.e., the sum of the
eigenvalues of ${{\bf{H}}_{\mathcal T}^H{{\bf{H}}_{\mathcal T}}}$ equals $|{\bf{H}}_{\mathcal T}|^2_{\rm F}$. Therefore, the norm-based low complexity ATAS scheme can be denoted as follows, where the active transmit antenna set is selected based on norm under given $M_a$ and then the search size can be reduced.

\begin{algorithm}\label{alg:2}\textbf{Norm-Based ATAS}
\begin{itemize}
\item \textbf{Initialization}: Set ${\zeta}_{\rm{temp}}= 0$. Sorting the columns of ${\bf{H}}$
as $|{\bf{H}}\left(:,\pi(1)\right)| \geq \ldots \geq |{\bf{H}}\left(:,\pi(M)\right)|$.
\item \textbf{For}: $M_a = 1 : M$
\begin{enumerate}
\item \textbf{Transmit antenna selection}: Choose ${\cal T}_{{M_a}} = \{\pi(1),\ldots,\pi(M_a)\}$, and the active channel matrix of $M_a$ selected transmit antennas is denoted as
${\bf{H}}_{{M_a}}$.
\item \textbf{Compute the EE}: Calculate the EE as $\zeta({\cal T}_{{M_a}})$ based on Algorithm \ref{alg:1}. If the solution candidate is infeasible, set $\zeta({\cal T}_{{M_a}}) = 0$.
\item \textbf{Compare the EE}: If ${\zeta}_{\rm{temp}} <
{\zeta}({\cal T}_{{M_a}})$, ${\zeta}_{\rm{temp}} = {\zeta}({\cal T}_{{M_a}})$, set ${\cal T} = {\cal T}_{{M_a}}$.
\end{enumerate}
\item \textbf{End For}
\item \textbf{If the solution candidate is infeasible, i.e., ${\zeta}_{\rm{temp}} = 0$, set ${\cal T} = \{1,\ldots,M\}$}.
\end{itemize}
\end{algorithm}

\subsection{Complexity Analysis}\label{sec4complex}

In this subsection, we provide complexity analysis for the proposed low complexity ATAS algorithm and the optimal exhaustive search. For the optimal exhaustive search, the search size is $\sum\nolimits_{j=1}^M C_j^M = \sum\nolimits_{j=1}^M \frac{M!}{j!(M-j)!} $, and thus the energy efficient iterative water-filling should be performed $\sum\nolimits_{j=1}^M \frac{M!}{j!(M-j)!}$ times, which increases exponentially as a function of the transmit antenna number $M$. For the proposed norm-based low complexity ATAS, the complexity of channel norm  sorting is negligible compared with the energy efficient iterative water-filling, and $M$ times of energy efficient iterative water-filling are required. Therefore, the complexity of the proposed norm based scheme is only $\frac{M}{\sum\nolimits_{j=1}^M \frac{M!}{j!(M-j)!}}$ of the complexity of optimal exhaustive search.

\subsection{Implementation Issue in Realistic Scenario}

During performing the ATAS, channel matrices of all antennas
are required for calculating the norm and determine the best active antennas. However, note that the inactive BS antennas should be switched off to save
energy. When the inactive antennas are switched off, the channel estimation related to these inactive antennas is impossible. Thus, there might exist time slots in which the channel matrices of the inactive BS antennas are not visible at the BS. When the channel matrices are invisible at the BS, the BS cannot utilize these information to perform ATAS, which would affect the EE performance. We refer this problem as {\emph{invisible CSIT problem}}.

In order to combat this drawback, a possible way is to add one dedicated training period to switch on all the BS antennas to help channel estimation. In this case, the power consumption of the training period would decrease the EE. Thus, the
above definition of EE serves as an upper bound. Nevertheless, there should be other low complexity schemes to combat the invisible CSIT problem.
For example, the statistical CSIT can be applied for the ATAS. The performance and cost tradeoff of these schemes is left for the future
work.


\section{Simulation Results}\label{Sec5}

We evaluate the performance under different scenarios to show
the effect of different system parameters. In the simulation,  pathloss and Rayleigh fading are considered. The
parameters are set based on \cite{Xu1}, where $W=5$MHz, the noise power is $-110$dBm, $P_{\rm{dyn}}=$83W, $P_{\rm{Sta}}$=45.5W, $\eta$=0.38, $P_{\max}$=46dBm and pathloss is $128.1+37.6 \log_{10}d$ with distance $d$ ($d$ in kilometers and all users are with the same distance). We use ``EE w Exh-AS'' to
denote the optimal energy efficient transmission with exhaustive
ATAS, ``EE w Norm-AS'' to denote the energy efficient
transmission with low complexity norm-based ATAS, ``EE wo AS'' to denote the energy efficient transmission
covariance optimization with activating all available BS antennas and 'SE' to denote the
spectral efficient transmission with activating all available BS
antennas and utilizing all available sum transmit power. Here, schemes with ``EE'' perform the transmit covariance optimization
determined according to energy efficient iterative water-filling and schemes with ``SE'' perform the transmit covariance optimization based on the spectral efficient iterative water-filling \cite{IterativeBC}.

\begin{figure}[t]
\begin{center}
\includegraphics[width=12.5cm] {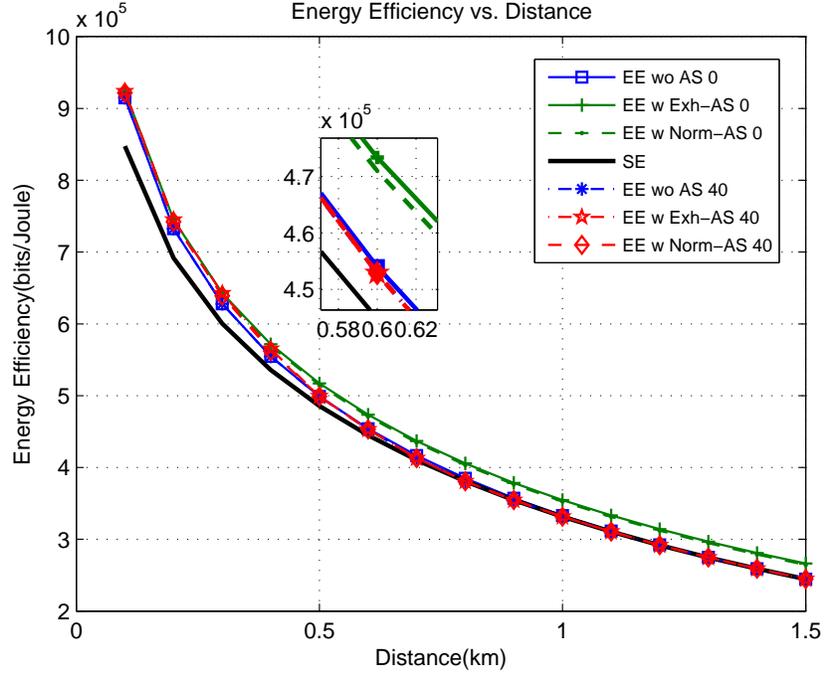}
\end{center}
\caption{EE versus distance, where $M=4, N=2, K=2, P_{\max} = 46$dBm.} \label{fig3}
\end{figure}

\begin{figure}[t]
\begin{center}
\includegraphics[width=12.5cm] {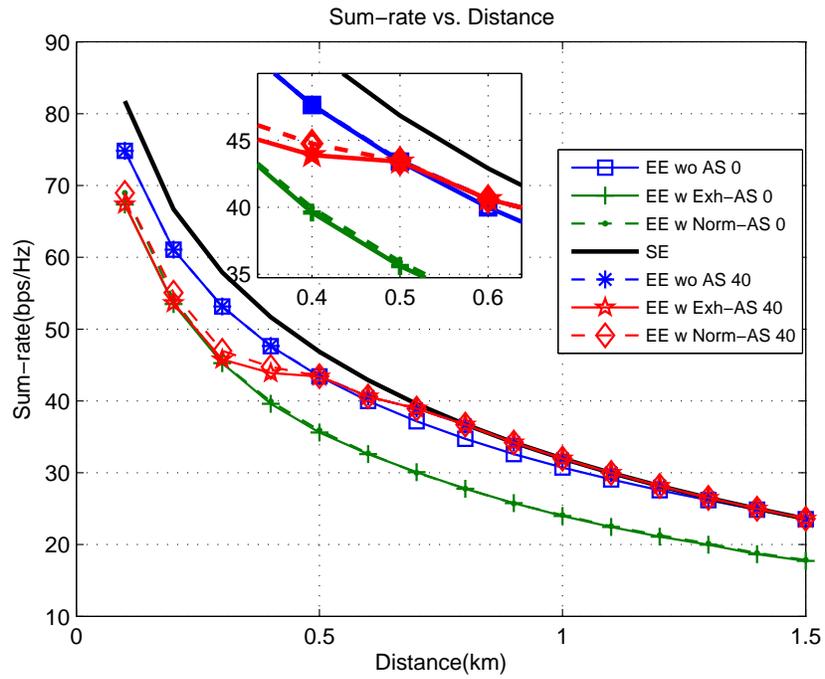}
\end{center}
\caption{Corresponding sum-rate versus distance, where $M=4, N=2, K=2, P_{\max} = 46$dBm.} \label{fig4}
\end{figure}

The EE and corresponding sum-rate versus pathloss are evaluated in Fig. \ref{fig3} and \ref{fig4} at first, where sum-rate constraints $0$ and $40$bps/Hz are both considered (sum-rate is bandwidth normalized as bps/Hz in the simulation). ``EE wo AS 0'' and ``EE wo AS 40'' denote the sum-rate constraints are 0 and $40$bps/Hz respectively, and so as the other schemes. In Fig. \ref{fig3},
we can see that schemes under ``EE w Exh-AS 0'' and ``EE w Norm-AS 0'' have best EE performance, where the gain comes from both the ATAS and energy efficient iterative water-filling. When the distance is short, e.g. 0.1km, the schemes with ``EE'' are all superior to ``SE''. As the distance increases, the EE of ``EE w Exh-AS 40'', ``EE w Norm-AS 40'' and ``EE wo AS 0'' degenerates into ``SE'' gradually. For ``EE w Exh-AS 40'' and ``EE w Norm-AS 40'', the degeneration comes from the minimum sum-rate requirement, since in a long distance scenario, the sum-rate would decrease significantly, so maximum transmit power should be utilized and all available transmit antennas should be activated to fulfill the minimum sum-rate requirement. For ``EE wo AS 0'', the degeneration is because that the globally optimal sum transmit power $P^*$ increases as the distance increases. When the the distance becomes significantly large, $P^*$ might be larger than $P_{\max}$, and then $P_{\max}$ should be utilized, i.e., ``EE wo AS 0'' degenerates into ``SE''. Look at Fig. \ref{fig4} then. We can see clearly that ``EE w Exh-AS 40'' and ``EE w Norm-AS 40'' coincide with ``SE'' when the sum-rate is smaller than 40bps/Hz, i.e., $d\geq 0.7$km, this explains the EE degeneration of the two schemes.

%

\begin{figure}[t]
\begin{center}
\includegraphics[width=12.5cm] {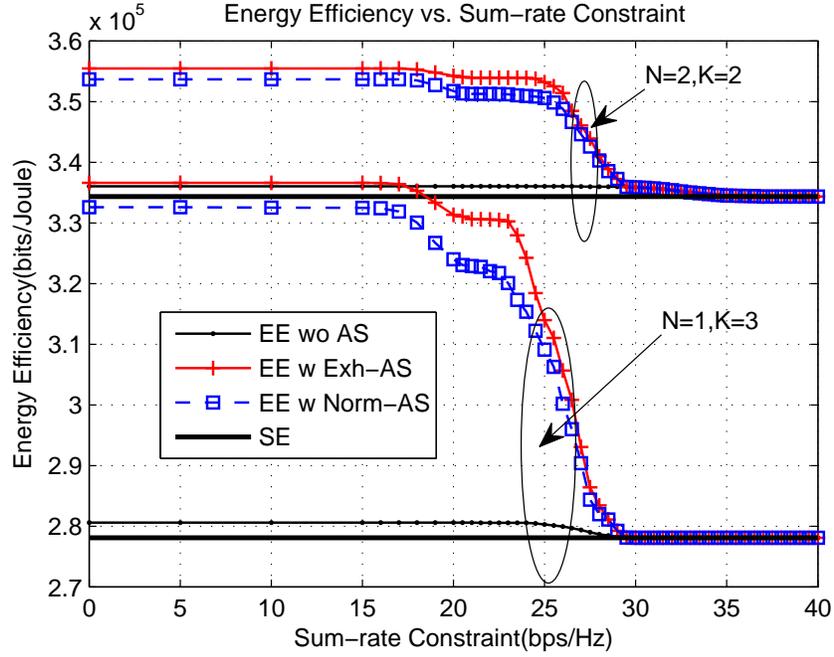}
\end{center}
\caption{EE under different sum-rate constraints, where $M=4, P_{\max} = 46$dBm, $d=$1km.} \label{fig003}
\end{figure}

\begin{figure}[t]
\begin{center}
\includegraphics[width=12.5cm] {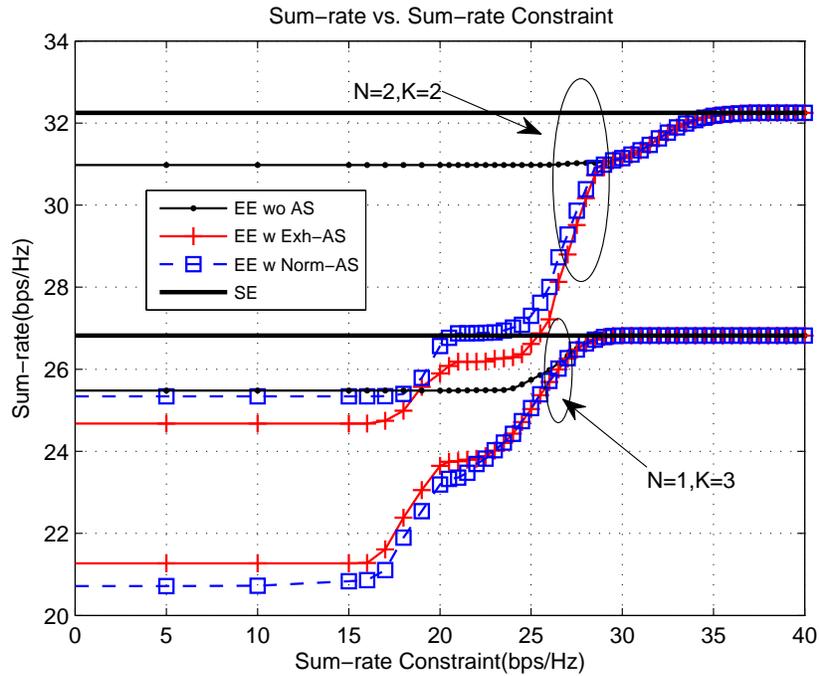}
\end{center}
\caption{Corresponding sum-rate under different sum-rate constraints, where $M=4, P_{\max} = 46$dBm, $d=$1km.} \label{fig0031}
\end{figure}

\begin{figure}[t]
\begin{center}
\includegraphics[width=12.5cm] {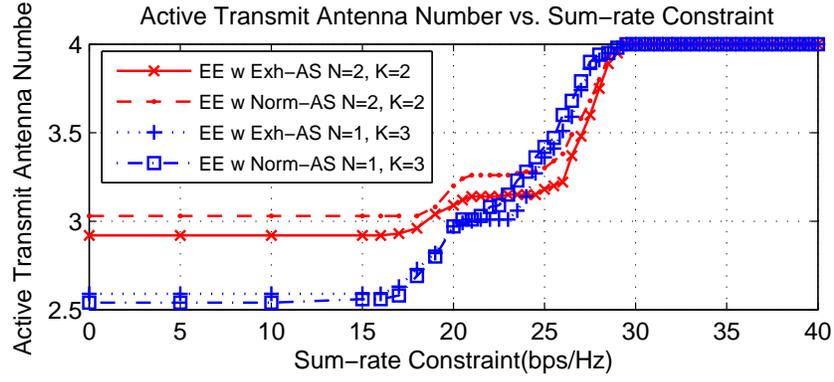}
\end{center}
\caption{Corresponding number of active transmit antennas  versus sum-rate constraints.} \label{fig00312}
\end{figure}

We simulate the effect of sum-rate constraints in Fig.
\ref{fig003}, \ref{fig0031}, \ref{fig00312}. We set $M=4, d=1$km, $P_{\max} = 46$dBm and consider
the case with $N=1,K=2$ and $N=1,K=3$. We can see that the ``EE w Exh-AS'' always
achieves the maximum EE, and ``EE w Norm-AS'' performs
very close to ``EE w Exh-AS''. Meanwhile, ``EE wo AS'' has smaller EE than ``EE w Norm-AS'', while ``SE'' has the worst EE. The
performance gain of ``EE'' with ATAS compared with ``EE wo AS'' comes
from the ATAS, as after active transmit antennas are
determined, turning off the inactive antennas can save the
dynamic power and then improve the EE. Meanwhile, the gap between
``EE wo AS'' and ``SE'' validates the efficiency of the energy efficient
transmit covariance optimization in section \ref{Sec3}.
The EE gap between different schemes are becoming smaller as the
sum-rate constraint increases. When the sum-rate constraint is
larger than 33(27) bps/Hz for $N=2,K=2$($N=1,K=3$), the four schemes perform the same. Correspondingly, looking at
Fig. \ref{fig0031}, the simulated sum-rate is fixed at 32.2(27) bps/Hz for $N=2,K=2$($N=1,K=3$),
and in this case, the maximum sum transmit power is utilized to
transmit{\footnote{In this case, the sum-rate constraint is
infeasible, and the maximum sum transmit power is employed.}}. Another observation is that the performance of  ``EE w Norm-AS'' and ``EE w Exh-AS'' both has a multi-stage feature. This feature can be explained according to Fig. \ref{fig00312}, i.e. number of active transmit antennas versus sum-rate constraints. In order to fulfill the increasing sum-rate constraint, the number of active transmit antennas  increases, and has a similar multi-stage feature. 

\begin{figure}[t]
\begin{center}
\includegraphics[width=12.5cm] {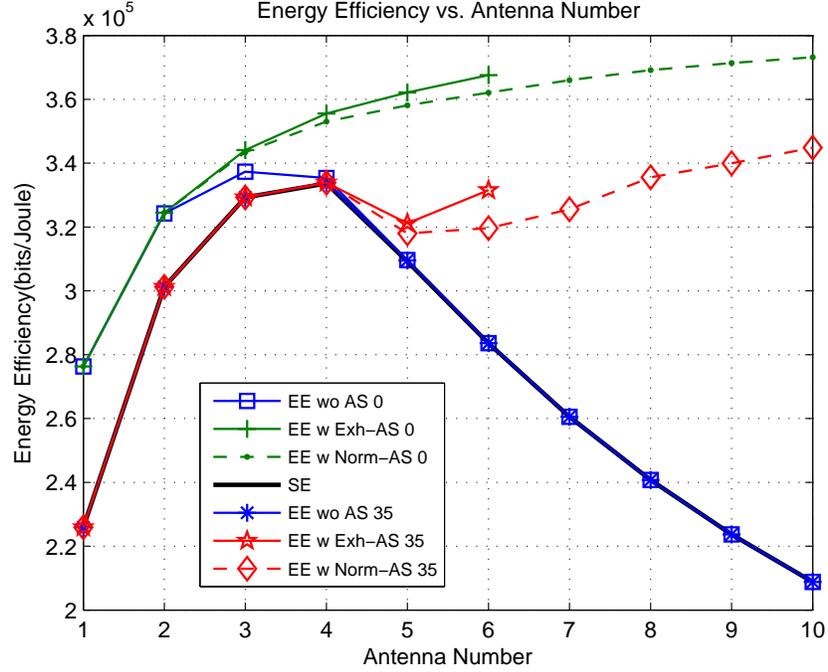}
\end{center}
\caption{EE under different BS antenna number, where $N=2, K=2, P_{\max} = 46{\rm{dBm}}$, $d=$0.5km.} \label{fig004}
\end{figure}

\begin{figure}[t]
\begin{center}
\includegraphics[width=12.5cm] {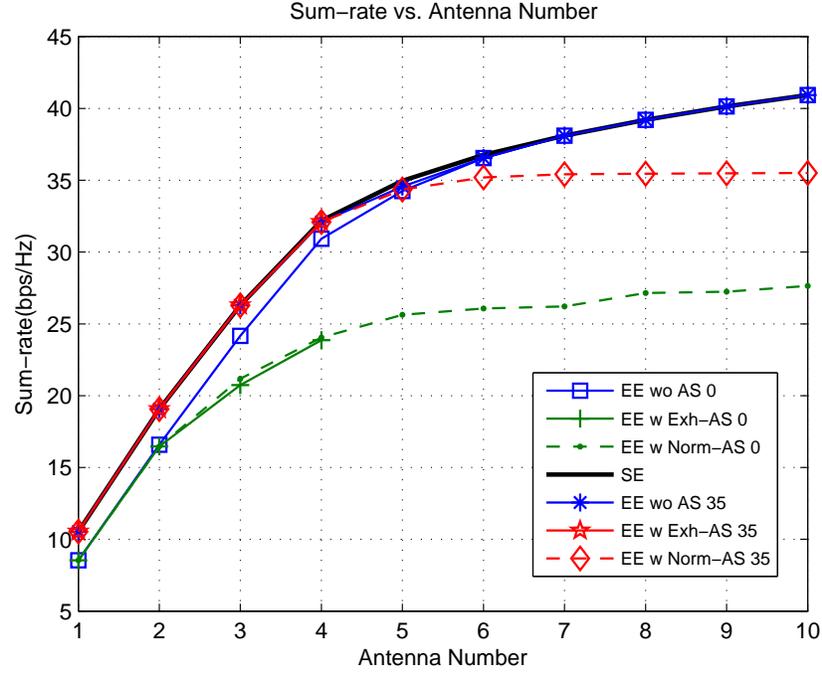}
\end{center}
\caption{Corresponding sum-rate under different BS antenna number, where $N=2, K=2, P_{\max} = 46{\rm{dBm}}$, $d=$0.5km.} \label{fig004:2}
\end{figure}

Fig. \ref{fig004} and \ref{fig004:2} depict the EE and corresponding sum-rate under different BS antenna configurations,
where $N=2, K=2, P_{\max} = 46{\rm{dBm}}, d=1$km. We consider that the sum-rate constraints are 0 and 35bps/Hz. Look at the case with sum-rate constraint 0 at first. ``EE w Norm-AS 0'' and ``EE w Exh-AS 0'' are both monotonously increasing as a function of the transmit antenna number $M$ at the BS. The performance gain comes from the transmit antenna selection diversity with suitable number of active transmit antennas  $M_a$. When the transmit antenna number increases, the probability of choosing channels of active antennas with better channel conditions increases. However, look at ``SE'' and ``EE wo AS 0''. The best EE performance is achieved when the antenna number is four and three respectively. As the spatial dimensions for DPC is $\min(N\times K,M)$  \cite{DPCScaling}, the multiplexing gains under $M=3$ and $M=4$ are three and four respectively, meanwhile, when $M=3$, there is also multiuser diversity, and thus the sum-rate under $M=3$ would be larger than $\frac{3}{4}$ of the sum-rate under $M=4$. Taking into account the practical power model, it is reasonable that achieved EE under different transmit covariance optimization techniques, i.e., SE and EE, is maximized at $M=3$ and $M=4$ respectively. Moreover, the behavior of ``EE wo AS 0'' can also be explained by Fig. \ref{fig00312}, where three active antennas are the optimal choice when the sum-rate constraint is 0 for $N=2,K=2$. The performance of ``SE'' and ``EE wo AS 0'' degenerates seriously when the antenna number is more than four. The reason can be explained by the multiplexing gain of the DPC. There are $\min(N\times K,M)$ spatial dimensions for DPC \cite{DPCScaling} and thus the multiplexing gain can scale as only $N\times K$ for the case with $M \geq N \times K$.
Meanwhile, the dynamic power increases linearly with $M_a$ in the power part, and then the EE loss with increasing dynamic power is significantly larger than the EE gain with sum-rate increasing when $M \ge 4$. Thus, the EE would decrease significantly for ``SE'' and ``EE wo AS 0''.
Let us look at the case with sum-rate constraint 35bps/Hz then. The behavior of ``EE w Norm-AS 35'', ``EE w Exh-AS 35'' can be understood according to the corresponding sum-rate in Fig. \ref{fig004:2}. When $M \leq 4$, all transmit antennas should be activated and maximum transmit power should be employed due to the sum-rate constraint, and thus ``EE w Norm-AS 35'' and ``EE w Exh-AS 35'' have the same EE as ``SE''. Since five active antennas can achieve the sum-rate of 35bps/Hz, and thus when $M > 5$, the EE of ``EE w Norm-AS 35'' is monotonously increasing as a function of $M$.


As shown above, more antenna number benefits for the EE with higher selection diversity under ATAS, however, it is worth emphasizing that configuring more antenna would cost higher Capital expenditures (CAPEX). In the design of the realistic systems, the tradeoff between the EE gain and the CAPEX loss should be taken into account.

\begin{figure}[t]
\begin{center}
\includegraphics[width=12.5cm] {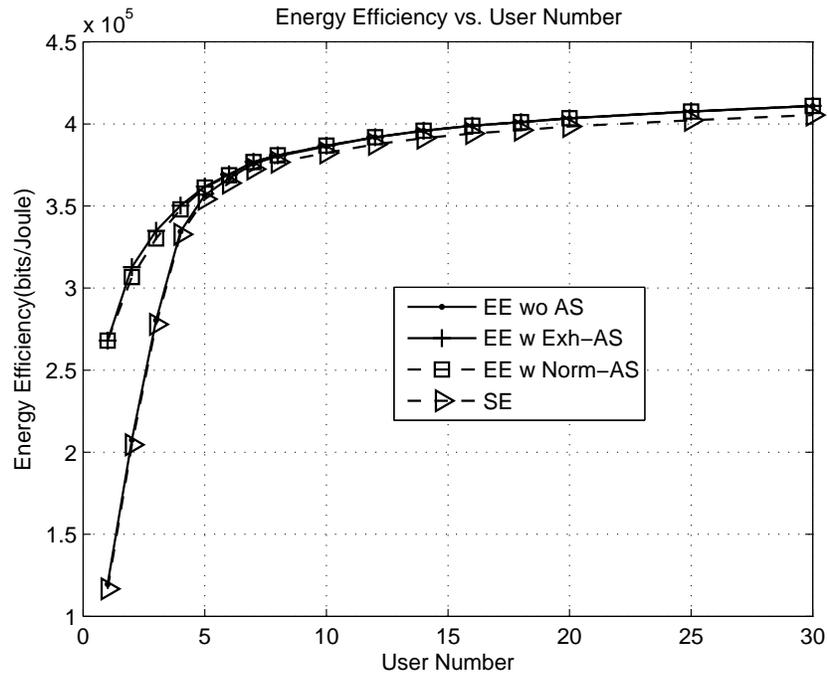}
\end{center}
\caption{EE under different user number, where $M=4, N=2, P_{\max} = 46{\rm{dBm}}$, and the sum-rate constraint is 0.} \label{fig005}
\end{figure}

We are interested in discussing the multiuser diversity finally through Fig. \ref{fig005}, which depicts the EE under different user number, where $M=4, N=2, P_{\max} = 46{\rm{dBm}}, C_{\min} = 0$, $d = 1$km. We can see that ``EE w Norm-AS'' and ``EE w Exh-AS'' degenerate into ``EE wo AS'', where all four transmit antennas should be active. Moreover, about the multiuser diversity, we can see from Fig. \ref{fig005} that a similar $M\log\log(NK)$ scaling law  can be acquired. Indeed, in our another work \cite{EEScaling}, we analyze the EE scaling law with the help of the Lambert $\omega$ function, and it is shown that when $ M_aP_{\rm dyn} + P_{\rm sta} > 0$, the multiuser diversity of  $\frac{M\log\log(NK)}{M_a P_{\rm dyn} + P_{\rm sta}}$ always holds. We can see that $M_a = M$ is optimal for the large user number case, which is distinct from the limited user number case.

\section{Conclusion}\label{Sec6}

We propose a novel optimization approach with transmit covariance optimization and ATAS to improve the EE in the MIMO-BC. Under a fixed active
transmit antenna set, we transform the EE of MIMO-BC based on uplink-downlink duality into a concave fractional  program, and then propose an energy efficient iterative water-filling scheme to maximize the EE for the MIMO-BC according to the block-coordinate
ascent algorithm. We prove the convergence of the proposed scheme and validate it through simulations. After determining the transmit covariance
under the fixed active transmit antenna set, we develop ATAS algorithms to further improve the EE, where exhaustive search and norm-based selection schemes are utilized. Through simulation results, the effect of system parameters on the EE is also discussed.

\section*{Acknowledgement}

The authors would like to thank the anonymous reviewers
for their valuable comments and suggestions to improve the
quality of the paper.

\bibliographystyle{IEEEtran}
\bibliography{reference}

\newpage

\newpage

\end{document}